\documentclass[conference,compsoc]{IEEEtran}
\IEEEoverridecommandlockouts
\usepackage{cite}
\usepackage{amsmath}
\usepackage{amssymb}
\usepackage{amsfonts}
\usepackage{amsthm}
\usepackage{algorithm}
\usepackage{algpseudocode}
\usepackage{graphicx}
\usepackage{textcomp}
\usepackage{xcolor}
\usepackage{soul}
\usepackage{enumitem}
\algnewcommand{\SectionComment}[2]{\Statex {\color{#2}$\blacktriangleright$ \textrm{#1}}}

\algnewcommand{\InlineComment}[1]{{\hspace{0.5em}\color{gray}\textrm{// #1}}}

\algnewcommand{\Phase}[1]{\SectionComment{#1}{jnSUDigitalRed}}

\MakeRobust{\Call}

\algrenewcommand\algorithmicwhile{\textbf{upon}}

\algrenewcommand\algorithmicloop{\textbf{wait}}

\algrenewcommand\algorithmicforall{\textbf{for each}}

\newcommand{\blackhalfcheckcross}{{\ding{51}}\rotatebox[origin=c]{-9.2}{\kern-0.7em{\ding{55}}}}

\definecolor{myblue}{HTML}{0072bd}
\definecolor{myorange}{HTML}{d95319}
\definecolor{mybrown}{HTML}{ad2c0a}

\definecolor{mygray}{gray}{0.6}
\definecolor{myred}{RGB}{192,50,26}
\definecolor{lightgray}{gray}{0.9}
\definecolor{mygreen}{RGB}{34,139,34}
\definecolor{transparentblue}{rgb}{0, 0, 1}
\colorlet{lightblue}{transparentblue!8}

\definecolor{jnSUDigitalRed}{HTML}{B1040E}
\definecolor{jnSUDigitalRedLight}{HTML}{E50808}
\definecolor{jnSUDigitalRedDark}{HTML}{820000}
\definecolor{jnSUDigitalBlue}{HTML}{006CB8}
\definecolor{jnSUDigitalBlueLight}{HTML}{6FC3FF}
\definecolor{jnSUDigitalBlueDark}{HTML}{00548f}
\definecolor{jnSUDigitalGreen}{HTML}{008566}
\definecolor{jnSUDigitalGreenLight}{HTML}{1AECBA}
\definecolor{jnSUDigitalGreenDark}{HTML}{006F54}

\def\BibTeX{{\rm B\kern-.05em{\sc i\kern-.025em b}\kern-.08em
    T\kern-.1667em\lower.7ex\hbox{E}\kern-.125emX}}
\usepackage{enumitem}
\setlist[enumerate]{itemsep=0mm}
\usepackage{stackengine}
\usepackage{bm}
\usepackage{tikz}
\usepackage{filecontents}
\usepackage{multirow}
\usepackage{booktabs}
\usepackage{color}
\usepackage{algorithm}  
\usepackage{algorithmicx}  
\usepackage{subcaption}
\usepackage{caption}
\usepackage[normalem]{ulem}
\usepackage{threeparttable}
\usepackage{pifont}
\usepackage{mathtools}
\usepackage{hyperref}
\usepackage{bbding}
\usepackage{textcomp}
\makeatletter

\newcommand{\Rmnum}[1]{\expandafter\@slowromancap\romannumeral #1@}
\makeatother

\definecolor{darkblue}{RGB}{48, 84, 151}
\definecolor{lightblue}{RGB}{222, 235, 246}
\definecolor{ao(english)}{rgb}{0.0, 0.5, 0.0}
\definecolor{applegreen}{rgb}{0.55, 0.71, 0.0}

\useunder{\uline}{\ul}{}

\newcommand{\tabincell}[2]{\begin{tabular}{@{}#1@{}}#2\end{tabular}}  

\usepackage[most]{tcolorbox}
\usepackage{xparse} 
\usepackage{lipsum}
\NewDocumentCommand{\answerbox}{m}{%
  \begin{tcolorbox}[
    colback=white,
    colframe=black,
    boxrule=0.5pt,
    sharp corners,
    left=3pt,
    right=3pt,
    top=3pt,
    bottom=3pt,
    fonttitle=\bfseries,
    coltitle=black,
    width=\linewidth,
  ]
  #1
  \end{tcolorbox}
}
\usepackage{fontawesome}  
\usepackage{tikz}

\newcommand{\progressbarA}[3]{%
  \begin{tikzpicture}[baseline]
    \node[anchor=east] at (0.07,0.100) {\makebox[0.2cm][r]{#3}}; 
    \ifnum#1>0 
      \fill[green!60!black] (0,0) rectangle (#1/#2*0.4,0.2); 
    \fi
    \fill[lightgray] (#1/#2*0.4,0) rectangle (0.4,0.2); 
  \end{tikzpicture}%
}

\newcommand{\progressbarFN}[3]{%
  \begin{tikzpicture}[baseline]
    \node[anchor=east] at (0.07,0.100) {\makebox[0.2cm][r]{#3}}; 
    \ifnum#1>0
      \fill[red!60!black] (0,0) rectangle (#1/#2*0.4,0.2); 
    \fi
    \fill[lightgray] (#1/#2*0.4,0) rectangle (0.4,0.2); 
  \end{tikzpicture}%
}

\newcommand{\progressbarTimeout}[3]{%
  \begin{tikzpicture}[baseline]
    \node[anchor=east] at (0.07,0.100) {\makebox[0.2cm][r]{#3}}; 
    \ifnum#1>0
      \fill[blue!60!black] (0,0) rectangle (#1/#2*0.4,0.2); 
    \fi
    \fill[lightgray] (#1/#2*0.4,0) rectangle (0.4,0.2); 
  \end{tikzpicture}%
}
\usepackage{pifont} 
\usepackage{siunitx}  

\usepackage[english]{babel}

\theoremstyle{definition}

\begin{document}


\title{Multi-Agent Collaborative  Fuzzing with Continuous Reflection \\for  Smart Contracts Vulnerability Detection
}




\author{
\IEEEauthorblockA {Jie Chen\textsuperscript{$*\dagger$}, Liangmin Wang\textsuperscript{$*\dagger$}}
\IEEEauthorblockA{
\textsuperscript{$*$}Southeast University, \textsuperscript{$\dagger$}Engineering Research Center of BASAM of Ministry of Education \\
}
}




\maketitle

\begin{abstract}
Fuzzing is a widely used technique for detecting vulnerabilities in smart contracts, which generates transaction sequences to explore the execution paths of smart contracts. 
However, existing fuzzers are falling short in detecting sophisticated vulnerabilities that require specific attack transaction sequences with proper inputs to trigger, as they i) prioritize code coverage over vulnerability discovery, wasting considerable
 effort on non-vulnerable code regions, and ii) lack semantic understanding of stateful contracts, generating numerous invalid transaction sequences that cannot pass runtime execution.


In this paper, we propose \texttt{SmartFuzz}, a novel collaborative reflective fuzzer  for smart contract vulnerability detection. It employs large language model-driven agents as the fuzzing engine and continuously improves itself by learning and reflecting through interactions with the environment.
Specifically, we firstly propose a new Continuous Reflection Process (CRP)  for fuzzing smart contracts, which reforms the transaction sequence generation as a self-evolving process through continuous reflection on feedback from the runtime environment.
Then, we present the Reactive Collaborative Chain (RCC)
to  orchestrate the fuzzing process into multiple sub-tasks based on the dependencies of transaction sequences.
Furthermore, we  design a multi-agent collaborative team, where  each expert agent is guided by the RCC to jointly generate and refine transaction sequences from both global and local perspectives.
We conduct extensive experiments to evaluate SmartFuzz’s performance on real-world contracts and DApp projects.
The  results  demonstrate that SmartFuzz outperforms existing state-of-the-art tools: (i)
it detects 5.8\%-74.7\% more vulnerabilities within 30 minutes. (ii) it reduces false negatives by up to 80\%. 
%



\end{abstract}




\section{Introduction}
Smart contracts are deployed in blockchain systems to enable the implementation of complex business logic in areas involving high-value assets,  with over one million transactions completed per day. Meanwhile, these smart contracts have also become alluring targets for potential attacks, which can lead to severe consequences  due to programming errors.
Specifically, attackers can illicitly steal cryptocurrencies or tokens by exploiting these bugs, as evidenced by several notorious  incidents, such as the TheDAO attack \cite{price2016digital} and the Parity Wallet Hack \cite{palladino2017parityhack}, which resulted in cumulative financial losses exceeding \$3.24 billion \cite{zhou2023sok}.
Smart contracts suffer from the inherent risk of being  prone to errors, owing to they are not patchable like traditional software, making it exceptionally challenging  to update or modify  contracts once they are deployed \cite{jiao2020semantic}.  
Thus, it is crucial to reveal vulnerabilities before deployment for ensuring the security of smart contracts.



\begin{figure}[!t]
    \centering
    \includegraphics[width=3.3in]{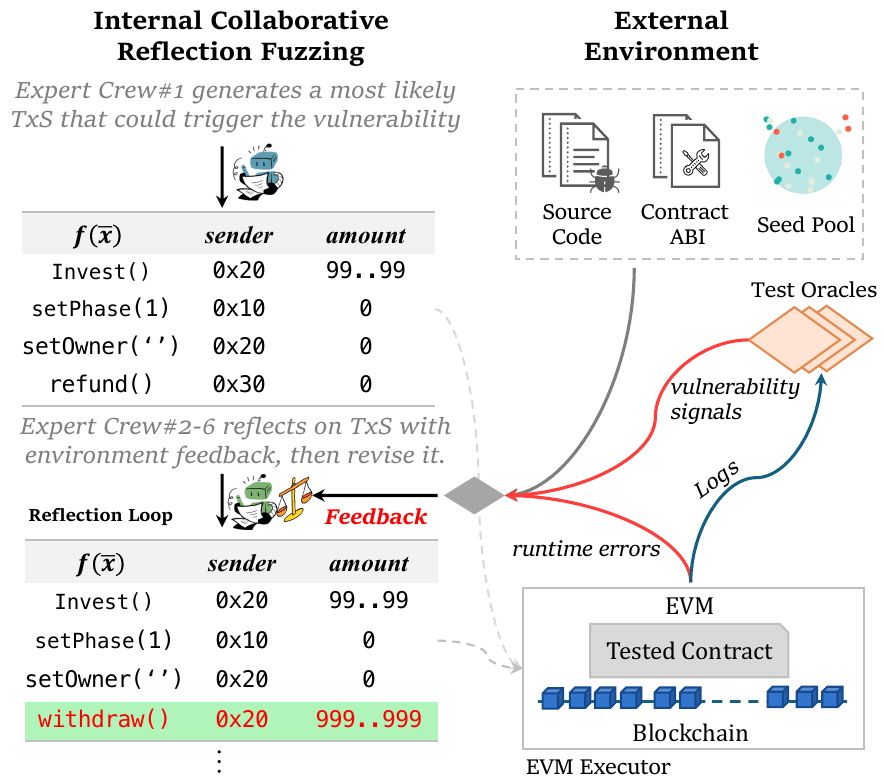}
    \caption{An Overview of SmartFuzz, which generates vulnerable (attack) transaction sequences that trigger the vulnerabilities. It continuously reflects and revises the generated  sequences with  external feedback during fuzzing process.}
    \label{fig:Overview}
\end{figure}

In recent years, pioneer researchers have developed a variety of tools to reveal vulnerabilities in smart contracts. Existing efforts can be roughly categorized into two groups. One branch of works \cite{luu2016making, mueller2017mythril, DBLP:conf/ndss/KalraGDS18, feist2019slither, bose2022sailfish} focus on static analysis (e.g., symbolic execution) to identify vulnerabilities. However, such approaches focus on static semantic and syntactic information, failing to consider the dynamic execution logic and runtime interactions that are critical for vulnerability detection. 
Another branch \cite{liu2018reguard, jiang2018contractfuzzer, nguyen2020sfuzz, he2019learning, DBLP:conf/kbse/SuDZZL22,qian2024mufuzz} 
adopts dynamic analysis techniques to uncover vulnerabilities in smart contracts.  Among these approaches, fuzzing has commonly been used to generate transactions as input to test smart contracts for vulnerability detection owing to its ability to 
produce a large number of test cases that explore diverse contract states.

Although fuzzing has achieved notable progress in detecting program vulnerabilities, it faces challenges due to the intractably large space of transaction combinations. Specifically, most fuzzers are designed to improve code coverage of smart contracts during the testing stage. However, simply increasing code coverage does not necessarily increase the number of vulnerabilities being found, which results in considerable resources being wasted on code regions that are unrelated to vulnerabilities \cite{bohme2022reliability}.
Besides, smart contracts are inherently stateful programs whose hidden vulnerabilities can only be triggered by specific vulnerable transaction sequences. Such sequences not only  require functions to be executed in a specific order, but also the input values for each function should be carefully crafted. However, existing fuzzers fail to understand program semantics in advance, resulting in the generation of numerous invalid transaction sequences that fail during  runtime execution.

To address the aforementioned challenges, we propose \texttt{SmartFuzz}, a novel  collaborative reflective fuzzing technique for smart contract vulnerability, as shown in Figure 1. It reforms the fuzzing process by leveraging large language model (LLM) agents and real-time environmental feedback to generate vulnerable transaction sequences.
Specifically, we propose a new Continuous Reflection Process (CRP) to model the transaction sequence generation as a self-evolving  process capable of interacting with the  environment during fuzzing process, such as the source code, application binary interface (ABI), seed pool and feedback from the Ethereum Virtual Machine (EVM) and test oracles.
Besides, we  design the Reactive Collaborative Chain (RCC), which decomposes the fuzzing process into multiple sub-tasks, thereby guiding SmartFuzz to capture global and local  dependencies of transaction sequences  during the CRP. Furthermore, we develop a multi-agent collaborative team, where each agent is assigned a specific sub-task and granted a set of authorized actions. Finally, we present the collaborative reflection policy, which instantiates the fuzzing policy with CRP and regulates the multi-agent team to follow the RCC in generating and refining vulnerable transaction sequences.
The experimental results demonstrate that SmartFuzz significantly outperforms existing tools across real-world  contracts and  DApp projects.

The main contributions of this paper can be summarized as follows:
\begin{itemize}
    \item We propose SmartFuzz, a novel collaborative reflective fuzzing technique for smart contract vulnerability detection. It employs LLM-driven agents as the fuzzing engine and continuously improves itself by  reflecting through interactions with the environment.

    \item   In SmartFuzz, we propose a new Continuous Reflection Process (CRP) that reforms the fuzzing process into a self-evolving process through continuous reflection on feedback from the runtime environment.

    \item We present the Reactive Collaborative Chain (RCC), which orchestrates the fuzzing process into multiple sub-tasks based on the dependencies of transaction sequences. We also design a multi-agent collaborative team, where  each expert agent is guided by the RCC to jointly generate and refine transaction sequences from both global and local perspectives.


    
    \item Extensive experimental results on real-world smart contracts and complex DApp projects demonstrate that SmartFuzz detects more true vulnerabilities than existing tools within a limited time.
\end{itemize}

\section{Motivation} 
\label{sec: motivation}

We first use a motivating example to illustrate why generating a proper transaction sequence is important for uncovering sophisticated vulnerabilities, and then discuss the challenges in smart contract fuzzing.

\begin{figure}[!t]
    \centering
    \includegraphics[width=3.3in]{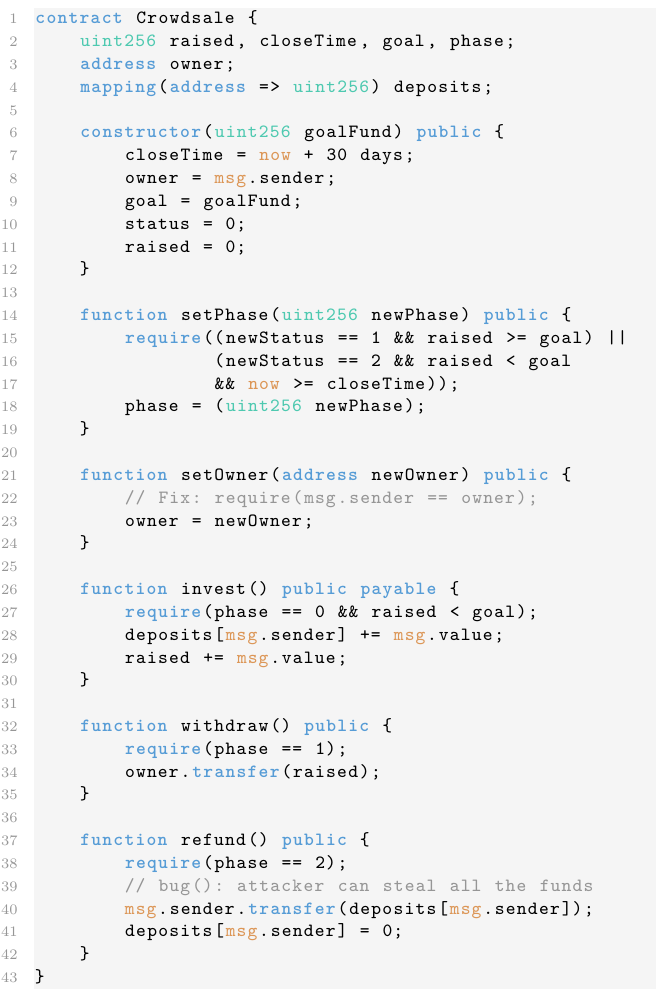}
    \caption{A Crowdsale contract (simplified version) containing vulnerability that allows attacker to steal funds.}
    \label{fig:CodeExample}
\end{figure}

\subsection{Motivating Example}

Inspired by RLF \cite{DBLP:conf/kbse/SuDZZL22} and ILF \cite{momeniMachineLearningModel2019}, we consider a Crowdsale contract as our motivating example, which is commonly used for time-bound fundraising in Ethereum to collect a certain amount of ethers.
Figure 1 illustrates a simplified version of the Solidity code of the Crowdsale contract. Firstly, the contract owner initializes the funding goal and the crowdsale duration (lines 7-9). Then, investors participate by calling the  \texttt{invest}  to contribute ethers during the active funding period. The contract maintains a status variable (\texttt{phase}) to indicate the current phase of the crowdsale. This status is updated to  success (\texttt{phase = 1}) if the collected funds meet or exceed the goal within the  period, or failure (\texttt{phase = 2}) if the deadline passes without reaching the target (line 18). Upon success, the owner is permitted to withdraw the raised funds by calling \texttt{withdraw} (lines 32-35); otherwise,  investors can call \texttt{refund} to claim refunds (lines 37-42).

This contract contains an Ether-Leaking vulnerability, which allows an attacker to steal the funds from the Crowdsale contract. Specifically, this vulnerability can be triggered by the following vulnerable transaction sequence:

\vspace{0.15em}
\begin{enumerate}[label=$t_{\arabic*}:$, itemsep=0pt, parsep=0pt, topsep=0pt]
  \item  Users call \texttt{invest} with \texttt{msg.value >= goal};
  \item An attacker with address \texttt{A} calls \texttt{setOwner(A)};
  \item The attacker calls \texttt{setPhase(newPhase)} \\ with \texttt{newPhase = 1};
  \item The attacker calls \texttt{withdraw} to steal the funds.
\end{enumerate}
\vspace{0.15em}
By executing such vulnerable transaction sequence, attackers can seize ownership and steal all funds without sending any ether.
 Given that a specific transaction sequence is necessary to trigger this vulnerability, most existing smart contract fuzzers may fail to generate the proper sequences to reveal this sophisticated vulnerability. 
The vulnerability arises from a missing pre-condition in the \texttt{setOwner} function, which must restrict access to \texttt{owner} (line 22). As a result, if there is a bug in the \texttt{refund} function (line 39), investors will be unable to claim their funds when the crowdsale fails.

\subsection{Challenges of Fuzzing Smart Contracts}

 \noindent\textbf{The Space of Transaction Combinations.} 
 Smart contracts are stateful programs  that maintains relevant state across transactions, such as the funds raised in our Crowdsale example (Figure 2).
 Their global states are altered by receiving and processing a series of transactions.  More formally, each transaction is identified by (i) a contract function $f$ to execute, together with any arguments $x$ required by the function, (ii) the address of the user $sender$ who initiates the transaction, and (iii) the amount of ether $amt$ transferred to the contract. Therefore, nontrivial bugs in smart contracts can only be reached after executing a specific sequence of transactions.  However, among the countless possible transaction sequences in smart contracts, it is difficult to find  the vulnerable transaction sequences. Each transaction in such sequences requires not only a specific order of functions but also proper input values for the  involved parameters.

\noindent\textbf{Limitation of Existing Fuzzers.} 
Unfortunately, it is challenging for existing fuzzers to tackle this challenge because of their limitations. For example,
Imitation Learning-based Fuzzer (ILF) \cite{he2019learning} leverages imitation learning from symbolic execution to explore deep execution paths, while Smartian \cite{choi2021smartian} combines static and dynamic analyses and mutates inputs according to data-flow coverage. Both these two fuzzers are coverage-guided fuzzers with the goal of maximizing the code coverage of smart contracts during fuzzing. 
However, in the case  of vulnerability detection, higher code coverage does not necessarily correlate with finding more vulnerabilities \cite{bohme2022reliability}. As indicated in \cite{durieux2020empirical}, only a few functions in smart contracts contain vulnerabilities, while most are benign, suggesting that blindly improving coverage is often inefficient for vulnerability discovery.
Besides, several fuzzers (e.g., ContractFuzzer \cite{jiang2018contractfuzzer}, sFuzz \cite{nguyen2020sfuzz}) concentrate on producing specific inputs that satisfy intricate intra-transaction conditions  (e.g., the statement require). However, these fuzzers ignore the order of transaction sequences, which is the key to revealing the complicated vulnerabilities as shown in the Crowdsale contract.
Furthermore, some fuzzers (Smartian \cite{choi2021smartian}, ConFuzzius \cite{torres2021confuzzius}, MuFuzz \cite{qian2024mufuzz})  rely on data dependency analysis to construct meaningful transaction sequences, but these approaches tend to reproduce sequences with normal  transaction behavior, instead of vulnerable transaction sequences 
 with specific function call orders and inputs that aim to trigger the underlying vulnerabilities.

\section{Methodology}
\label{sec: Methodology}

To address the aforementioned challenges, we propose \texttt{SmartFuzz}, a novel  collaborative reflective fuzzer for smart contract vulnerability. 
We first  model the transaction sequence generation as a new proposed continuous reflection process (CRP). Then, we present the Reactive Collaborative Chain to decomposes the  fuzzing process into multiple  sub-tasks to facilitate real-time hierarchical reflection with the external environment. Furthermore, we develop a multi-agent collaborative team to generate vulnerable transaction sequences that can trigger underlying vulnerabilities.

\subsection{Continuous Reflection Process}
\label{sec: Continuous Reflection Process}

We define the Continuous Reflection Process as a  new process to model smart contract fuzzing process of the transaction sequence generation. 
The key insight behind CRP is two-fold: 1) Smart contract fuzzing process need specific transaction sequences with proper inputs to adequately explore the program execution paths for digging out the underlying vulnerabilities; 2) LLMs are often fail that output broken transaction sequences that cannot pass the backend execution, so they should  continuously correct their answers based on the feedback.

In this process, the LLM agents iteratively  revise previous  state by conducting global and local reflection on real-time environmental feedback. 
Formally, an CRP is defined as a tuple as follows:
\begin{equation}
  CRP = (\mathcal{S}, \mathcal{A}, \mathcal{T}, \mathcal{E}, \mathcal{F})
\end{equation}

\noindent where $\mathcal{S}$ is a set of states, $\mathcal{A}$ is a set of revise actions of agents after reflection and $\mathcal{T}: \mathcal{S} \times \mathcal{A} \rightarrow \mathcal{S}$ is the state transition process. Besides,  $\mathcal{F}$ is the real-time feedback when external environment $\mathcal{E}$   executing current state. 

At each reflection step $i$, the multi-agent team $\mathcal{G}$, guided by a collaborative policy $\pi$, observes the current state $s_i$ and executes a series of actions $A_i$. Specifically, each agent $g \in \mathcal{G}$ will incorporate feedback as an inherent element of its prompt. The revise actions $\mathcal{A}: \mathcal{S} \times \mathcal{F} \rightarrow A$ are chosen based on the previous state $S_{i-1}$ and the feedback ${F}(S_{i-1})$ from the environment. This action choosing process of each agent can be  defined as follows:
\begin{equation}
  a_i^g \sim \mathcal{R}^{g_{k} \in  \mathcal{G}}_{\ i}(S_{i-1}, \mathcal{F}(S_{i-1}, a_{i-1}), \pi), i \in [1, n] 
\end{equation}
\noindent where $\mathcal{R}^{g \in  \mathcal{G}}_{\ i}$ represents the reflection operation of agent $g_{k}$ at step $i$. 
Then, all the actions from each agent are aggregated to form a set of revise actions $A_i$:
\begin{equation}
  A_i = \{a_i^{g_{k}} \mid g_k \in \mathcal{G}\}_{k=1}^m
\end{equation}
where $m$ is the number of agents in the team $\mathcal{G}$.
Besides, the state transitions to  $s_{i+1}$  based on the aggregated actions, computed as:
\begin{equation}
  S_{i+1} = \mathcal{T}(S_i, A_i) 
\end{equation}

The process repeats iteratively until the maximum reflection rounds are reached, or the feedback contains a stop signal (e.g., a vulnerability being detected). Our Goal is to find  vulnerable transaction sequences that can trigger underlying vulnerabilities by continuously reflecting upon and revising prior failed instances.

\begin{table}[!t]
  \centering
  \caption{Mapping from CRP concepts to fuzzing concepts.}\label{tab: CRP}
  \resizebox{0.47\textwidth}{!}{ 
      \begin{tabular}{ll}
      \toprule
      CRP Concept & Fuzzing Concept  \\  
      \midrule
      State $s \in \mathcal{S}$ & Previous Transaction Sequence $T$ \\  
      Revise Action $a \in \mathcal{A}$ & Transaction mutation \\  
      Transition $\mathcal{E}$ & Sequence updating $\mathcal{T} \rightarrow \mathcal{T}^{\prime}$ \\  
      Environment feedback $\mathcal{F}$ & Feedback from EVM and test Oracles \\  
      Collaborative policy $\pi$ & Policy for generating $T$ \\  
      Reflection Operation $\mathcal{R}$ & Test case generation \\  
      Multi-agent team $G$ & Fuzzer with policy $\pi$ \\  
      \bottomrule
      \end{tabular}
  }
\label{table:1}
\end{table}


\begin{figure*}[!t]
  \centering
  \includegraphics[width=6.8in]{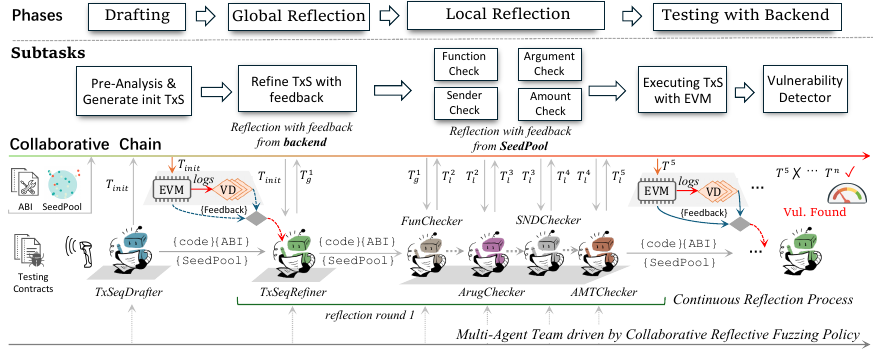}
  \caption{An workflow of SmartFuzz. It decomposes smart contract fuzzing into four phases, each further divided into specific subtasks via a collaborative chain. The designed multi-agent team operates under a collaborative reflective policy (instantiated by CRP) to generate transaction sequences that trigger underlying vulnerabilities.}
  \label{fig:team}
\end{figure*}


\subsection{Modeling Fuzzing as CRP} 
\label{sec: Modeling Fuzzing as CRP}

Table 1 shows how the key components of CRP connect to concepts in smart contract fuzzing. 
Unlike prior works \cite{he2019learning, DBLP:conf/kbse/SuDZZL22} that model smart contract fuzzing as a Markov decision process, where each state represents a single transaction and generates each transaction iteratively, we represent each state  $s \in \mathcal{S}$  as a complete transaction sequence $T = (t_1, t_2, \dots, t_n)$. This allows the fuzzing policy to generate an entire transaction sequence at once.
At step  $i$, based on the previous sequence $T_{i-1} = (\bar{t}_1, \bar{t}_2, \ldots, \bar{t}_n)$ and the block state trace $b_{\text{init}} \xrightarrow{\bar{t}_1} \cdots \xrightarrow{\bar{t}_n} b_n$ resulting from executing  $T_{i-1}$  on the contract  $c $ under test, the agent reflects on  $T_{i-1}$  using environmental feedback  $\mathcal{F}(T_{i-1})$  while following a  policy $\pi$. Specifically, the reflection operation  $\mathcal{R}^{i-1}$ from policy $\pi$  produces a series of actions $A_{i-1}$ to revise the $T_{i-1}$ and then output next state transaction sequence $T_{i}$.
Then, $T_{i}$ will be executed, updating the block state trace from the scratch, and the agents receive a new feedback $\mathcal{F}(T_{i})$, such as runtime errors and vulnerability signals.

\noindent\textbf{Transaction Sequence Generation.} Following prior works \cite{he2019learning, DBLP:conf/kbse/SuDZZL22}, a transaction sequence contains multiple transactions, each of which can be defined as a generation process based on previous transactions $\bar{t}$:
\begin{equation}
  \bar{t} = (f(x), sender, amt)
\end{equation}
where  $f_i$  is a callable function selected from the contract; 	$x_i$  are the input parameters generated for  $f_i$; $sender_i$  is the assigned sender address; $amt_i$  is the transaction amount.

 We then model the basic fuzzing policy of transaction sequence generation as a continuous reflection process, and it can be captured by the function $\pi : T_{i} \times A_i \rightarrow T_{i+1}$, which is designed to  output a new transaction sequence based on the current transaction sequence, feedback-driven actions.
Since smart contract transactions have complex structure, we decompose this  basic fuzzing policy into four components:
\begin{enumerate}
    \item[1)] $\pi_{func} : T_f  \times A  \times F \rightarrow : T_{f}^*$, used to select a \textit{function} $f$ with action $A$ for each transaction from $F = \{f^1, \ldots, f^n\}$, the set of (public or external) functions of the tested contract.
    
    \item[2)] $\pi_{args} : T_{args} \times A \times F \times  \mathcal{X} \rightarrow T_{args}^*$, used to select  \textit{arguments} $x$ with action $A$ for the selected function $f$, where $\mathcal{X}$ is the set of possible function argument values.
    
    \item[3)] $\pi_{sender} : T_{sender}^*  \times A \times \textit{SND} \rightarrow T_{sender}^*$, used to select a \textit{sender} from a pre-defined set of possible senders $\textit{SND}$.
    
    \item[4)] $\pi_{amt}: T_{amt} \times A \times  F \times \textit{AMT} \rightarrow T_{amt}^*$, used to select \textit{payment amount} with action $A$ for the selected function $f$, where $\textit{AMT}$ is the set of possible ether amounts that can be sent.
\end{enumerate}

Consequently, each component of a transaction is generated by a set of distinct sub-policies. Specifically, the functions in the transaction sequence of the next state are generated by $\pi_{func}(\bar{t})$; their corresponding arguments are sampled by $\pi_{args}(\bar{t}, f)$; the sender addresses are selected by $\pi_{sender}(\bar{t}, f)$, and the transaction amounts are chosen by $\pi_{amt}(\bar{t}, f)$.
Notably, we further  instantiate $\pi$ as a collaborative reflection policy $\pi^{cr}$ to enable adaptive and goal-directed fuzzing with a multi-agent team (detailed in § 4.3.2).

\noindent\textbf{Interaction with External Environment}.
As depicted in right part of Figure 1, we define the external environment  as $\mathcal{E} = (Input, SeedPool, EVM, Detector)$. The input component includes the  source code of testing contracts, and corresponding ABI files. The \textit{SeedPool} maintains seed-level information, including sender addresses and transaction amounts, which are used to construct candidate transactions. 
The \textit{EVM} acts as the execution engine that commits the generated transactions and processes them  according to required specifications, while the \textit{Detector} inspects the  execution trace logs to identify potential vulnerabilities.
Specifically, EVM and vulnerability detector are regarded as the backend of the external environment $\mathcal{E}$, which provides real-time feedback $\mathcal{F}$ (e.g., runtime errors and vulnerability reports) to guide the fuzzing process.


\subsection{The SmartFuzz System} 
\label{sec: The SmartFuzz System}

As shown in Figure 3, we  present the SmartFuzz system, which includes three main components: 1) a reactive collaborative chain decomposes transaction sequence generation into multiple subtasks and orchestrates agent  different agents for these subtasks to build a feasible workflow; 2) 
a multi-agent collaborative team, where each agent functions as a specific expert, performs feedback-driven  actions through interaction with the external environment. 3) a collaborative reflection policy that guides the multi-agent team to iteratively  revise transaction sequences by conducting hierarchical reflection  on real-time environmental feedback.

\noindent\textbf{Reactive Collaborative Chain.}
Finding specific vulnerable transaction sequences is challenging due to the intractably large search space arising from the combinatorial explosion of possible transactions.
Existing fuzzers \cite{he2019learning, DBLP:conf/kbse/SuDZZL22} generate transactions one by one without the ability to plan the entire sequence in advance, often failing to reach the depth needed to trigger vulnerabilities. 
Therefore, we design the Reactive Collaborative Chain (RCC) to guide the generation of transaction sequences, by considering the sequence dependencies from both global and local perspectives.
This allows SmartFuzz to avoid getting trapped in local optima and to produce a complete transaction sequence at each step. 

Specifically, RCC decomposes the task of generating a complete transaction sequence into four distinct phases: \textit{drafting}, \textit{global reflection}, \textit{local reflection}, and \textit{testing}. As shown in Figure 3, each phase is further broken down into  manageable subtasks that are assigned to LLM agents with specialized expertise. 
In a multi-agent system, 
the failures of any individual agent
can cascade through the system and invalidate the entire task \cite{cemri2025multi}.
To mitigate such risks, RCC adopts a dependency scheduling strategy that  prioritizes global reflection before local reflection, ensuring that the high-level structure of the transaction sequence is semantically coherent before engaging in fine-grained refinements. Besides, within local reflection, \texttt{function check} precedes other subtasks, as it dictates the permissible argument types and value transfers required by each transaction. Notably, each agent in the collaborative team has the access to the program context module, which includes the source code of testing contracts, corresponding Application Binary Interface (ABI) files and SeedPool.

RCC guides each crew agent in the collaborative team to focus on specific subtasks to iteratively refine the transaction sequences through multi-turn interactive feedback from  environment. 
The subtasks in each phase can be defined as:
\begin{equation}
\begin{aligned}
\mathcal{C} = \langle \mathcal{P}^1, \mathcal{P}^2, \dots, \mathcal{P}^{\lvert\mathcal{C}\rvert} \rangle \\
\mathcal{P}^i = \langle  \tau ^1_i,  \tau ^2_i, \dots,  \tau _{i}^{\lvert \mathcal{P}^i\rvert} \rangle \\
\tau ^j =   \langle \mathbb{I}, g^j, P^j \rangle 
\end{aligned}
\end{equation}

\noindent where transaction sequence generation task $\mathcal{C}$ includes a series of phases $\mathcal{P}^i$, and each phase has several subtasks (e.g., $\tau^j$ is the $j$-th subtask in phase $i$). Each subtask $\tau^j$ consists of an input $\mathbb{I}$ (program context), an agent $g^j$, and required profile $P^j$ of agent.
Hence, the entire process of transaction sequence generation  along the reactive collabrative chain can be formulated as:
\begin{equation}
  \begin{aligned}
  (\tau^0, \tau^1 \leadsto \mathcal{D}) \xrightarrow{\mathcal{F}} (\tau^2 \leadsto \mathcal{R}_{\ g})_\circlearrowright \xrightarrow{\mathcal{F}} (\tau^3 \dots \tau^7 \leadsto \mathcal{R}_{\ l})_\circlearrowright
  \end{aligned}
\end{equation}

\noindent where $\mathcal{D}$ denotes the drafting phase, $\mathcal{R}_g$ the global reflection phase, and $\mathcal{R}_l^n$ the local reflection phase. The symbol $\leadsto$ indicates the specific agents responsible for executing the subtasks within each phase. The term ${\mathcal{F}}$ represents the feedback obtained from the external environment, which is used to guide subsequent reflection phases. The symbol $\circlearrowright$ indicates that the corresponding subtask participates in a reflection loop, where the agent iteratively refines its output based on  feedback from different step.

\noindent\textbf{Multi-Agent Collaborative Team.}
%
This section presents the design of the multi-agent collaborative team. For each agent we specify its profile, goals, and the actions it can execute to complete its assigned subtask along the collaborative chain.

\noindent\textit{Profiles for Agents.} Each agent is assigned a specific role with tailored capabilities to address different aspects of the transaction sequence generation process.
The collaborative team consists of six specialized agents with distinct goals: \texttt{TxSeqDrafter} generates initial transaction sequences by analyzing contract code and identifying potentially vulnerable functions; \texttt{TxSeqRefiner} improves sequences based on execution feedback; \texttt{FunChecker} validates each function in the generated transaction sequences to ensure it is  callable; \texttt{ArgChecker} ensures argument correctness; \texttt{SNDChecker} and \texttt{AMTChecker} verify  $amount$ and $amount$ in each transaction to ensure they originate from the seed pool.

\noindent\textit{Environmental Feedback Extraction.}  The feedback extracted from backend are categorized into two main types: runtime feedback and vulnerability feedback.
Specifically, runtime feedback is derived from error signals produced by the EVM when executing invalid or exception-triggering transaction sequences. In contrast,  vulnerability feedback, is derived from the  reports produced by the testing oracle, which examines the execution  logs from EVM to detect whether any vulnerability is triggered by the transaction sequence.

Raw information from these two backend modules  are often low-level, redundant, or semantically ambiguous, making them difficult for the LLM agents to understand. Therefore, 
we define $S_{EVM}$ as the runtime error signals, focusing on errors related to key transaction elements $(f, x, sender, amount)$, and $S_{Vul}$ as the signals from the test oracle indicating vulnerability detection results. SmartFuzz then automatically translates these signals ($S_{EVM}, S_{Vul}$) into pre-defined feedback ($F_{EVM}, F_{Vul}$) to guide the continuous reflection process. For example, \texttt{FunctionNotFound} indicates the target contract lacks the specified function, \texttt{ArgumentMismatch} indicates a mismatch in the number or types of function arguments, \texttt{SenderError} covers issues like insufficient funds or invalid nonce, \texttt{NonPayableFunction} indicates sending Ether to a non-payable function and \texttt{IncorrectTransactionValue} means an amount inconsistent with the value constraints defined by the payable functions. Besides, vulnerability feedback is considered as \texttt{VulnerabilityFound(type)} and \texttt{VulnerabilityNotFound}. The former indicates a specific vulnerability  is detected (e.g., \texttt{VulnerabilityFound(EL)} for Ether Leaking), while the latter means no vulnerabilities are found.


\noindent\textit{Permission-Aware Actions.} 
To prevent inter-agent conflicts, we further design permission-aware actions, which constrain each agent, assigned to a specific  sub-task, from inadvertently modifying elements of the transaction sequence that are beyond its responsibility. All actions for each agent are detailed in Table 6.
For example, \texttt{TxSeqDrafter} focuses on actions such as \texttt{findFuncs} to identify callable functions and \texttt{pickVulFuncs} to select potentially vulnerable functions, while \texttt{ArgChecker} specializes in \texttt{checkTypes} to ensure type correctness of arguments. Each action is implemented through carefully designed prompts that guide the LLM's responses toward specific outputs, creating a structured workflow that enables complex collaborative behaviors to emerge from relatively simple individual capabilities.

\noindent\textbf{Collaborative Reflective Policy.}
We further instantiate the collaborative reflective policy $\pi^{cr}$ with the CRP as the fuzzing policy, which is with tailored architecture of multi-agents collaborative team.  \texttt{SmartFuzz} leverages the collaborative reflection policy $\pi^{cr}$ to detect vulnerabilities by continuously refining transaction sequences through multi-agent collaboration. This policy is built upon the preliminary three phases defined within the reactive collaborative chain:

\textit{Drafting $\mathcal{D}$}: This phase comprises two subtasks (pre-analysis $\tau^0$ and generate the initial transaction sequence $\tau^2$), which are performed by the \texttt{TxSeqDrafter} agent $g^{1}$ with its own profile $P^{1}$.   We treat  a set of input  (e.g., source code $c$, corresponding ABI  file $c_{abi}$, seed pool $S_{pool}$) as context information. The generation of the initial transaction sequence $T_0$ is represented as follows:
\begin{equation}
     \{c, c_{abi}, S_{pool}\} \cup (\tau^1, \tau^2)\xrightarrow{g^{1}(P^{1})\rightsquigarrow A^1} T_0
\end{equation}
\noindent where the \texttt{TxSeqDrafter} agent $g^{1}$ analyzes the input information, and performs a series of permission-aware actions $A^1$ based on the task requirements. Specifically, it first identifies the set of callable functions $F_{call}$ along with their required parameters (line 1). It then autonomously selects relevant functions to form a function sequence $(f_1, \cdots, f_n, f_i \in F_{call})$ and, for each function $f_i$, generates appropriate input parameters $\bar{x}_i$, chooses sender addresses $sender_i$, and specifies transaction amounts $amount_i$ from $S_{pool}$ (line 2). The agent further compiles these into a complete sequence of transactions $T_0$, which serves as the starting point for the following  phases.

\textit{Hierarchical Reflection.} In Global Reflection ($\mathcal{R}_{g}$), SmartFuzz refines a previous transaction sequence as a whole by considering its overall structure and dependencies (e.g., the input sequence is $T_0$ in the first reflection round). Specifically,  global reflection relies on feedback from the external environment to guide sequence-level improvements. This process can be represented as follows:
\begin{equation}
  \hspace*{-0.5em}
    \begin{aligned}
    (t_{0}, \cdots, t_{n}) \cup \left\{c, c_{abi}, S_{pool}\right\} \cup \left\{F_{EVM}, F_{Vul} \right\}  \cup (\tau^3) \\ \xrightarrow{g^{2}(P^2)\rightsquigarrow A_2} (t_{0}^{\prime}, \cdots, t_{n}^{\prime}) 
    \end{aligned}
\end{equation}

\noindent where  $t_0, \cdots, t_n$ denote the previously generated transaction sequence, $g^1$ refers to the agent \texttt{TxSeqRefiner}, and $F_{EVM}$ and $F_{Vul}$ represent feedback  from the EVM and the vulnerability detector, respectively—the latter identifies vulnerabilities based on test oracles. Besides,  $t_{0}^{\prime}, \cdots, t_{n}^{\prime}$ denotes the new  sequence produced after  global reflection.

In local Reflection ($\mathcal{R}_{l}$), SmartFuzz   focuses on element-wise check for each transaction within the sequences. This phase is designed to alleviate hallucinations in LLMs by preventing syntactically plausible yet semantically incorrect or irrelevant outputs from accumulating.
Specifically, function names are first  validated to ensure they are indeed callable within the smart contract, rather than being indiscriminately generated by large language models. This validation serves to safeguard against erroneous outputs that could compromise the contract's functionality, which can be expressed as:
\begin{equation}
  \hspace*{-0.3em}
    (f_{0}, \cdots, f_{i}) \cup \left\{c, c_{abi}\right\}  \cup (\tau^4)\xrightarrow{g^{3}(P^3)\rightsquigarrow A_3} (f_{0}^{\prime}, \cdots, f_{i}^{\prime}) 
\end{equation}
\noindent where $f_{0}, \cdots, f_{i}$ are the function names in the previous transaction sequence, and $g^3$ represents the agent \texttt{FunChecker}. This agent checks whether each function is callable within the smart contract, ensuring that only valid functions are included in the transaction sequence.
Furthermore, to ensure that the remaining three local elements are both valid and capable of guiding the fuzzing process toward vulnerable program branches, we assign the agents \texttt{ArgChecker} $g_4$, \texttt{SNDChecker} $g_5$, and \texttt{AMTChecker} $g_6$ to handle their respective sub-tasks $\tau$. Specifically, each agent first evaluates whether the current value of its assigned element is valid and likely to contribute to vulnerability discovery. If not, \texttt{ArgChecker} generates a new argument, while \texttt{SNDChecker} and \texttt{AMTChecker} select alternative sender addresses and transaction amounts from the SeedPool. In either case, the agent prioritizes values that increase the likelihood of triggering potential vulnerabilities. These processes can be defined as follows:

\begin{equation}
    \begin{aligned}
    (\bar{x}_{0}, \cdots, \bar{x}_{i}) \cup \left\{c, c_{abi}\right\} \cup (f_{0}^{\prime}, \cdots, f_{i}^{\prime}) \cup (\tau^5) \\ \xrightarrow{g^{4}(P^4)\rightsquigarrow A_4} (\bar{x}_{0}^{\prime}, \cdots, \bar{x}_{i}^{\prime})
    \end{aligned}
\end{equation}

\begin{equation}
    \begin{aligned}
        (sender_0, \cdots, sender_{i}) \cup \left\{c, c_{abi}, S_{pool}\right\} \\ 
        \xrightarrow{g^{5}(P^5) \rightsquigarrow A_5} (sender_0^{\prime}, \cdots, sender_{i}^{\prime})
    \end{aligned}
\end{equation}
\begin{equation}
    \begin{aligned}
    (amt_0, \cdots, amt_i) \cup \left\{c, c_{abi}, S_{pool}\right\} \cup (f_{0}^{\prime}, \cdots, f_{i}^{\prime}) \\ \xrightarrow{g^{6}(P^6)\rightsquigarrow A_6} (amt_0^{\prime}, \cdots, amt_i^{\prime}) 
   \end{aligned}
\end{equation}

\noindent where $\bar{x}_{0}, \cdots, \bar{x}_{i}$ are the argument values in the previous transaction sequence, $sender_0, \cdots, sender_{i}$ are the sender addresses, and $amt_0, \cdots, amt_i$ are the transaction amounts. As a result, in each iteration of the reflection loop, \texttt{SmartFuzz} refines the transaction sequence based on environmental feedback, ensuring that both the global structure and the local elements are validated and progressively optimized to increase the likelihood of triggering vulnerabilities.

\subsection{Vulnerability Detection}
\label{sec: Vulnerability Detection}

To detect vulnerabilities, inspired by \cite{he2019learning, choi2021smartian,DBLP:conf/kbse/SuDZZL22, qian2024mufuzz}, we set up eight test oracles, which are listed in Table 2 with corresponding brief descriptions. Specifically, we extract the execution log (e.g., executed opcodes) of the tested smart contract during fuzzing and apply the test oracles to identify whether the vulnerabilities are triggered.

\section{Evalution}
\label{sec:exps}

In this section, we conduct an extensive experimental evaluation of
SmartFuzz by addressing the following questions:

\begin{itemize}
    \item \textbf{RQ1:} Does SmartFuzz detect vulnerabilities more  effectively than
existing state-of-the-art tools? 
    \item \textbf{RQ2:} How does the continuous reflection process impact SmartFuzz's performance, and how do different reflection rounds affect vulnerability detection across various LLM engines?
    \item \textbf{RQ3:}  How does SmartFuzz perform on real-world contracts to detect a broader spectrum of vulnerabilities?

\end{itemize}

\subsection{Experimental Setup}
\label{sec:setup}

\noindent\textbf{Implementation of SmartFuzz.}
We run SmartFuzz on a modified backend
based on ILF \cite{he2019learning}, RLF \cite{DBLP:conf/kbse/SuDZZL22} that supports fast executing transactions natively without performing Remote Procedure Call (RPC). We implement our LLM  engines with Ollama \footnote{https://ollama.com/} on a local server. Given our server’s limited computational power, we employ open-source LLMs with 8B or 16B parameters, such as CodeGemma, CodeLlama, LLaMA3, CodeQwen, and DeepSeek-R1.
Commercial closed-source large models, such as ChatGPT and Claude, are excluded because of their prohibitively expensive API costs. Additionally, we employ the CrewAI \footnote{https://www.crewai.com/} framework to build the multi-agent collaborative system. The source code of the SmartFuzz is available at

\noindent\textbf{Datasets.} 
We construct two datasets to evaluate the performance of SmartFuzz and baselines. The first dataset, denoted as D1, is derived from a labeled dataset \cite{so2021smartest}  containing contracts with Ether-Leaking and Suicidal Contract vulnerabilities, both of which typically require specific transaction sequences to be triggered. Since different tools support varying vulnerability  scopes, we select only the 85 vulnerable contracts that can be successfully analyzed by all the evaluated tools. Among these contracts, 108 functions contain Ether-Leaking vulnerabilities and 46 functions exhibit Suicidal Contract vulnerabilities. 
Although D1 includes only two types of vulnerabilities, we adopt this dataset to ensure comparability with prior closely related studies, which follow the same experimental settings.
To further evaluate detection performance across a broader range of vulnerabilities in real-world contracts, we adopt the dataset aggregated by Mufuzz \cite{qian2024mufuzz} as our second dataset, referred to as D2. From this dataset, we choose six additional categories of vulnerabilities, comprising a total of 108 labeled vulnerable contracts.  Besides, we collect 34  DApp projects from popular platforms (e.g., GitHub and Etherscan) to evaluate the performance of SmartFuzz in practical scenarios.  

\noindent\textbf{Comparing Baselines and Configurations.}
We primarily consider existing tools that are publicly available and can support transaction generation. These tools can be  categorized into symbolic execution (e.g., Mythril \cite{mueller2017mythril}, SmarTest \cite{so2021smartest}) and fuzzing (e.g., ILF \cite{he2019learning}, RLF \cite{DBLP:conf/kbse/SuDZZL22},   Smartian \cite{choi2021smartian}, Echidna \cite{echidna2019}, sFuzz \cite{nguyen2020sfuzz}, ConFuzzius \cite{torres2021confuzzius}, Mufuzz \cite{qian2024mufuzz}). Given varying functionalities and supported vulnerability types across tools, we strategically select tailored baselines for each research question to ensure fair comparisons. Specifically, for RQ1, we use Mythril, SmarTest, Smartian, ILF, and RLF as baseline tools. For RQ2, we analyze SmartFuzz's performance using different LLM engines, such as CodeGemma, CodeLlama, LLaMA3, CodeQwen, and DeepSeek-R1.
For RQ3, we select Oyente, Mythril, sFuzz, ConFuzzius, Smartian, and Mufuzz to evaluate their detection performance on a broader range of vulnerabilities in real-world contracts.  We follow prior studies (ILF \cite{he2019learning}, RLF \cite{DBLP:conf/kbse/SuDZZL22}, Mufuzz \cite{qian2024mufuzz}) in experimental settings for all these tools.
Moreover, all experiments are conducted on a server with two Intel Xeon E5-2678 v3 CPUs (24 cores and 48 threads), 128GB of RAM, and four 2080Ti GPUs.



\subsection{Effectiveness of SmartFuzz (RQ1)}
\label{sec:results}

Following prior works that support transaction sequence generation \cite{so2021smartest, he2019learning, DBLP:conf/kbse/SuDZZL22}, we evaluate the effectiveness of SmartFuzz in detecting Ether-Leaking and Suicidal Contract vulnerabilities. We Compared SmartFuzz with existing state-of-art tools  on D1 dataset, including Mythril, SmarTest, Smartian, ILF and RLF. 
To ensure a fair comparison, we configured the analysis timeout for SmartFuzz to 30 minutes, consistent with the settings applied to Mythril and SmarTest in RLF and ILF. For RLF, a reinforcement learning-based method, we set the reward  to 0.7, consistent with its original settings. 
Furthermore, to align with the standard 2-fold cross-validation strategy employed in the baseline methods, we also randomly split the D1
 dataset into two partitions. We independently evaluated SmartFuzz on each partition to obtain experimental results across the entire dataset.

 Following these settings, we then identify the reported vulnerabilities according to the labels in D1. Additionally, Figure 4 illustrates the number of true vulnerabilities detected by various tools over time (in seconds).  Within a 30-minute duration (1,800 seconds), SmartFuzz showcases its efficiency in vulnerability detection, finding the most number of true vulnerabilities. We next compare SmartFuzz with other symbolic executors and fuzzers.

 \begin{table}[!t]
  \centering
  \caption{Vulnerabilities reported by different tools in D1. Numbers in parentheses indicate the number of functions with vulnerabilities. TP: true positives, \#: total reported vulnerabilities. EL refers to Ether-Leaking vulnerabilities, SC refers to Suicidal Contract vulnerabilities.}
  \label{table:Effectiveness}
  \resizebox{0.45\textwidth}{!}{
  \begin{tabular}{llcccccc}
  \toprule
  \multirow{2}{*}{\textbf{Type}} & \multirow{2}{*}{\textbf{Tool}} & \multicolumn{2}{c}{\textbf{EL (108)}} & \multicolumn{2}{c}{\textbf{SC (46)}} & \multicolumn{2}{c}{\textbf{ALL (154)}} \\
  \cmidrule(lr){3-4} \cmidrule(lr){5-6} \cmidrule(lr){7-8}
  & & \textbf{TP} & \textbf{\#} & \textbf{TP} & \textbf{\#} & \textbf{TP} & \textbf{\#} \\
  \midrule
  
  \multirow{2}{*}{\shortstack[l]{Symbolic \\ Executors}}
  & Mythril       & 9  & 12  & 26 & 26 & 35 & 38 \\
  & SmarTest      & 62 & 63  & 41 & 41 & 103 & 104 \\
  \midrule
  
  \multirow{4}{*}{Fuzzers} 
  & Smartian      & 21 & 21  & 22 & 22 & 43 & 43 \\
  & ILF           & 86 & 88  & 43 & 43 & 129 & 131 \\
  & RLF           & 98 & 100 & 43 & 43 & 141 & 143 \\
  & \textbf{SmartFuzz}  & \textbf{105} & \textbf{105} & \textbf{45} & \textbf{45} & \textbf{150} & \textbf{150} \\
  \bottomrule
  \end{tabular}}
  \end{table}

\noindent\textbf{Comparing with Existing Symbolic Executors.} 
As shown in Table 2, SmartFuzz detects 150 true vulnerabilities, outperforming Mythril by 74.7\% \(\left(\frac{150 - 35}{154}\right)\) and RLF by 5.8\% \(\left(\frac{150 - 141}{154}\right)\). Specifically, SmartFuzz identifies 105 EL and 45 SC, exceeding Mythril's results by 88.0\% \(\left(\frac{105 - 9}{108}\right)\) and 41.3\% \(\left(\frac{45 - 26}{46}\right)\), respectively.
Both SmartFuzz and SmarTest report substantially more true vulnerabilities than Mythril, as the latter wastes considerable effort on covering non-vulnerable code for merely high code coverage. Meanwhile, SmartFuzz finds 39.8\% \(\left(\frac{105 - 62}{108}\right)\) more EL and 8.7\% \(\left(\frac{45 - 41}{46}\right)\) more SC than SmarTest.
This improvement stems from SmartFuzz’s ability to generate transaction sequences involving multiple interacting functions, which are often required to trigger complex vulnerabilities. However, SmarTest struggles with such multi-function cases (e.g., the Ether-leaking example in Figure 1).
Additionally, Figure 4 shows that SmartFuzz detects most vulnerabilities within the first 600s, significantly faster than other baselines. In contrast, SmarTest exhibits limited performance during the initial 600s, but the number of its detected vulnerabilities increases rapidly in the subsequent 900s (from 900s to 1,800s). This suggests that its performance is constrained by the low efficiency of the symbolic execution mechanism (e.g., the SMT solver).

\noindent\textbf{Comparing  with Existing Fuzzers.} As shown in Table 2, SmartFuzz outperforms all the fuzzers in terms of the total number of true vulnerabilities detected. For example, SmartFuzz detects 69.48\% ($\frac{150-43}{154}$), 13.64\% ($\frac{150-129}{154}$), 5.84\% ($\frac{150-141}{154}$) more true vulnerabilities than Smartian, ILF and RLF separately. 
This is mainly because ILF is a coverage-guided fuzzer with an aim to cover as many execution paths as possible, inevitably leading to huge time consumption. Meanwhile, ILF and RLF treat the generation of transaction sequence as a  markov decision process, which means the next transaction is generated only based on the previous transaction. This process would collapse  when the previous generated transaction is inappropriate. In contrast, SmartFuzz generates transaction sequences in a collaborative reflection way, which can generate complete transaction sequences and continuously correct each transaction within every sequence.

\answerbox{
\textbf{Answer to RQ1:} SmartFuzz is more efficient in detecting Ether-Leaking (EL) and Suicidal (SC) vulnerabilities than existing tools. Within 30 minutes, SmartFuzz detects 5.8\%-74.7\% more vulnerabilities than  baselines.
}

\begin{figure}[!t]
  \centering
  \includegraphics[width=3.3in]{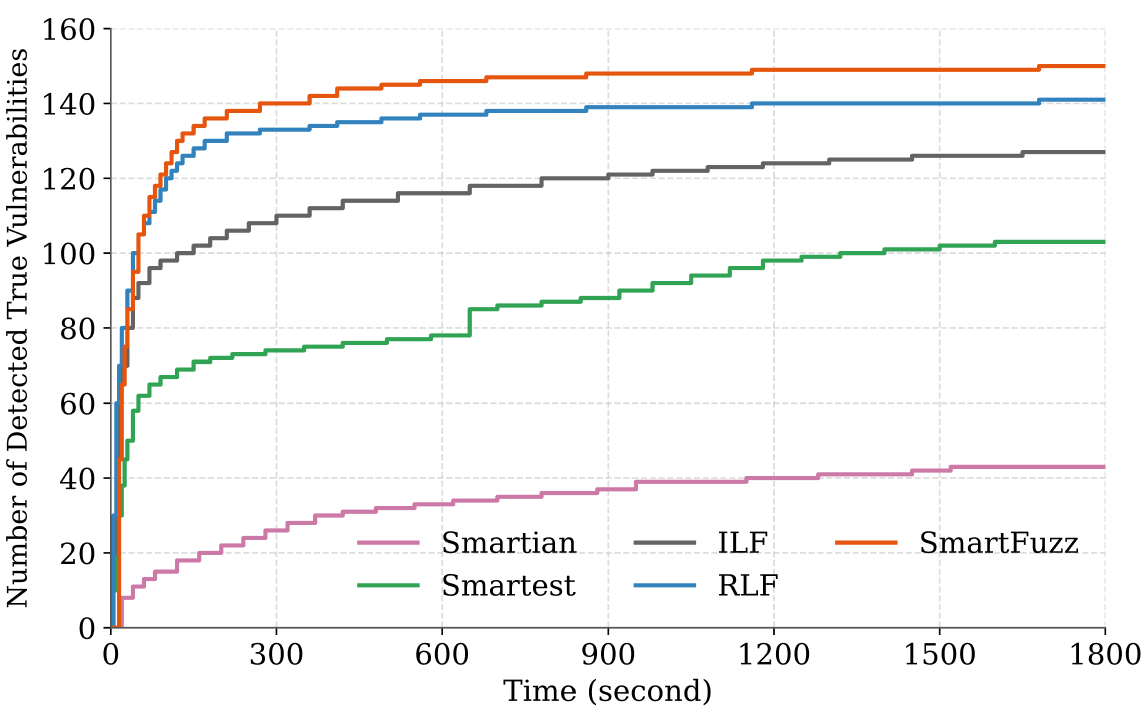}
  \caption{True vulnerabilities being found versus time (second) by different tools in D1.}
  \label{fig:timeVulNumD1}
\end{figure}

\begin{figure}[!t]
  \centering
  \includegraphics[width=3.3in]{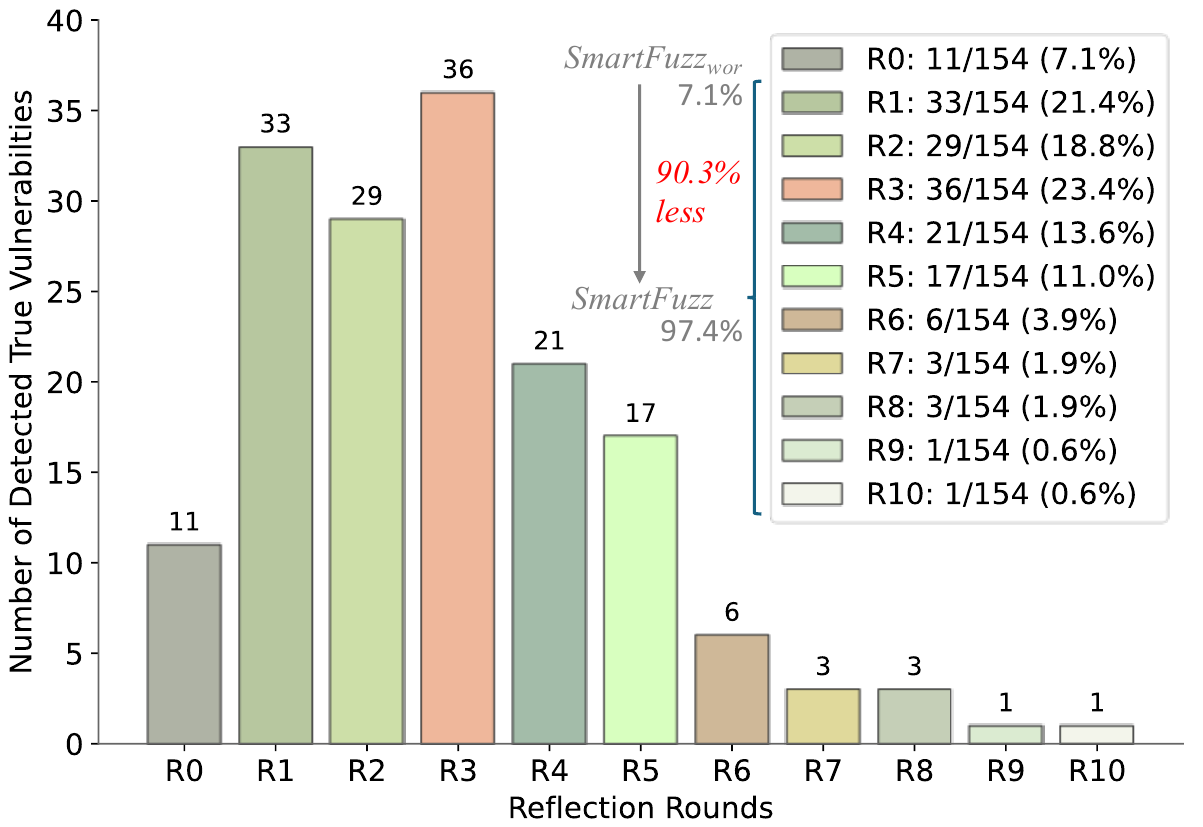}
  \caption{Ablation results on the importance of the continuous reflection process. To remove  this process, we set the $max\_reflection\_round$ to 0. For the rest of the reflection rounds, we set the $max\_reflection\_round$ to 10.}
  \label{fig:ReflectionAblation}
\end{figure}

\subsection{Ablation Study (RQ2)}
\label{sec:ablation}

In this section, we evaluates the contribution of the continuous reflection process to SmartFuzz’s performance and further investigates the relationship between different reflection rounds and the number of true vulnerabilities detected. Besides, we also investigate how different  LLM engines influence its performance.

\begin{figure*}[!t]
  \centering
  \includegraphics[width=7.0in]{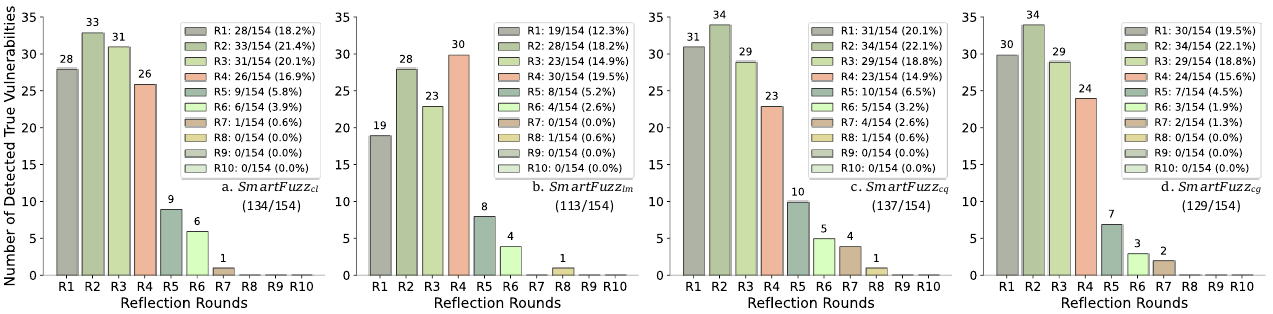}
  \caption{Ablation Study on investigating the relationship between reflection round and the number of detected vulnerabilities when using across different large language models. }
  \label{fig:differentLLM}
\end{figure*}

\noindent\textbf{Effectiveness of Continuous Reflection Process.} 
As shown in Figure 5, we set $max\_reflection\_round$ to 0 in SmartFuzz$_{wor}$ to remove the impact of the reflection process. 
Furthermore, we run SmartFuzz with $max\_reflection\_round = 10$ and track the exact number of reflection rounds required to detect each vulnerability, in order to explore the relationship between reflection round and the number of true vulnerabilities detected.


Specifically, with DeepSeek-R1 as the default LLM engine,
experimental results show that SmartFuzz$_{wor}$ suffers a significant 90.3\% performance drop compared to SmartFuzz, with the number of detected true vulnerabilities decreasing from 150 to  11. This highlights the critical role of the continuous reflection process in SmartFuzz, and further demonstrates that relying solely on large language models is insufficient for effectively discovering true vulnerabilities.
Notably, the number of vulnerabilities discovered peaks at the three reflection rounds, demonstrating that as the number of reflection rounds increases, the fuzzing process benefits from the accumulation of previously failed attempts. This process effectively guides the generation of more targeted  transaction sequences, enabling the detection of vulnerabilities with more complex triggering conditions.
Furthermore,  88.3\% of all vulnerabilities are uncovered within the  five reflection rounds, highlighting the importance of continued reflection in identifying edge-case vulnerabilities that may be missed in early stages.

\noindent\textbf{Impact of Different LLM Engines.} To explore the impact of varying LLM engines on the performance of vulnerability detection, we evaluate SmartFuzz with four additional models. Since leading commercial models like ChatGPT and Claude are not open-source and our GPU resources are limited, we focus on widely adopted open-source models, including CodeLLaMA (7b), LLaMA3.1 (8b), CodeQwen (7b), and CodeGemma (7b). 
We instantiate SmartFuzz with these LLM engines, denoted as SmartFuzz$_{cl}$, SmartFuzz$_{lm}$, SmartFuzz$_{cq}$, and SmartFuzz$_{cg}$, respectively.
Notably, 
most LLMs offer multiple versions with different parameter sizes. Although we selected models with parameter sizes as close as possible, exact alignment across all models is inevitably inconsistent.

Figure 6 shows the relationship between reflection rounds and the number of true vulnerabilities detected by SmartFuzz with different LLM engines. 
Specifically, the experimental results reveal that SmartFuzz$_{cl}$, SmartFuzz$_{cq}$, and SmartFuzz$_{cg}$ achieve comparable performance, detecting 
87.0\%,  89.0\%, and 83.8\% of all vulnerabilities, respectively. 
All three  outperform SmartFuzz$_{lm}$, identifying 10.4\% to 15.6\% more true vulnerabilities, largely because the LLM engines behind them are specifically optimized for program code, benefiting their ability to  understand program semantics. Moreover, we find that SmartFuzz${cl}$, SmartFuzz${cq}$, and SmartFuzz$_{cg}$ suffer from a marked rise in hallucinations when the reflection depth exceeds five rounds.
This substantially undermines their ability to detect vulnerabilities with complex triggering conditions, leading to inferior performance relative to the default SmartFuzz.

\answerbox{
\textbf{Answer to RQ2:} Our results demonstrate three key findings: (1) the continuous reflection process is critical, with SmartFuzz suffering a 90.3\% performance drop when reflection is disabled; (2) extended reflection rounds progressively uncover more complex vulnerabilities, with 88.3\% detected within five rounds; and (3) code-specialized LLMs outperform general models by detecting 10.4\%-15.6\% more vulnerabilities.
}




%

\begin{table*}[!t]
  \centering
  \caption{The performance of SmartFuzz on real-world contracts. The number of true positives (TP), false negatives (FN), and timeout or error (TE) cases  raised by each tool. False negatives indicate that a tool fails to detect the corresponding bugs, and `n/a' denotes that a tool does not support detecting the bug class. Besides, contracts from the DApp Projects are marked as fail if their execution time exceeds 10 minutes or the maximum number of reflection rounds.}
  \label{tab:tool-comparison}
  {
   \setlength{\tabcolsep}{1.45pt} 
  \resizebox{1.0\textwidth}{!}{ 
  \begin{tabular}{l|*3{c}|*3{c}|*3{c}|*3{c}|*3{c}|*3{c}|*3{c}}
  \toprule
  \multirow{3}{*}{\bfseries Types} & \multicolumn{6}{c|}{\bfseries Symbolic Executors} & \multicolumn{15}{c}{\bfseries Fuzzers} \\
  \cmidrule(lr){2-7} \cmidrule(lr){8-22}
   & \multicolumn{3}{c|}{Oyente} & \multicolumn{3}{c|}{Mythril} & \multicolumn{3}{c|}{sFuzz} & \multicolumn{3}{c|}{ConFuzzius} & \multicolumn{3}{c|}{Smartian} & \multicolumn{3}{c|}{MuFuzz} & \multicolumn{3}{c}{\bfseries SmartFuzz} \\
   &  TP &  FN &  TE &  TP &  FN &  TE &  TP &  FN &  TE &  TP &  FN &  TE &  TP &  FN &  TE &  TP &  FN &  TE &  TP &  FN &  TE \\
  \midrule
  \textbf{BD} (20) & \progressbarA{5}{20}{5} & \progressbarFN{14}{20}{14} & \progressbarTimeout{1}{20}{1} & \progressbarA{8}{20}{8} & \progressbarFN{5}{20}{5} & \progressbarTimeout{7}{20}{7} & \progressbarA{10}{20}{10} & \progressbarFN{10}{20}{10} & \progressbarTimeout{0}{20}{0} & \progressbarA{10}{20}{10} & \progressbarFN{4}{20}{4} & \progressbarTimeout{6}{20}{6} & \progressbarA{11}{20}{11} & \progressbarFN{9}{20}{9} & \progressbarTimeout{0}{20}{0} & \progressbarA{15}{20}{15} & \progressbarFN{5}{20}{5} & \progressbarTimeout{0}{20}{0} & \progressbarA{19}{20}{19} & \progressbarFN{1}{20}{1} & \progressbarTimeout{0}{20}{0} \\
  \textbf{UE} (31) & \multicolumn{3}{c|}{n/a} & \progressbarA{10}{31}{10} & \progressbarFN{3}{31}{3} & \progressbarTimeout{18}{31}{18} & \progressbarA{12}{31}{12} & \progressbarFN{19}{31}{19} & \progressbarTimeout{0}{31}{0} & \progressbarA{16}{31}{16} & \progressbarFN{13}{31}{13} & \progressbarTimeout{2}{31}{2} & \progressbarA{19}{31}{19} & \progressbarFN{12}{31}{12} & \progressbarTimeout{0}{31}{0} & \progressbarA{27}{31}{27} & \progressbarFN{4}{31}{4} & \progressbarTimeout{0}{31}{0} & \progressbarA{31}{31}{31} & \progressbarFN{0}{31}{0} & \progressbarTimeout{0}{31}{0} \\
  \textbf{UD} (17) & \multicolumn{3}{c|}{n/a} & \progressbarA{5}{17}{5} & \progressbarFN{6}{17}{6} & \progressbarTimeout{6}{17}{6} & \progressbarA{10}{17}{10} & \progressbarFN{7}{17}{7} & \progressbarTimeout{0}{17}{0} & \progressbarA{10}{17}{10} & \progressbarFN{7}{17}{7} & \progressbarTimeout{0}{17}{0} & \progressbarA{2}{17}{2} & \progressbarFN{15}{17}{15} & \progressbarTimeout{0}{17}{0} & \progressbarA{17}{17}{17} & \progressbarFN{0}{17}{0} & \progressbarTimeout{0}{17}{0} & \progressbarA{17}{17}{17} & \progressbarFN{0}{17}{0} & \progressbarTimeout{0}{17}{0} \\
  \textbf{EF} (22) & \multicolumn{3}{c|}{n/a} & \multicolumn{3}{c|}{n/a} & \progressbarA{7}{22}{7} & \progressbarFN{15}{22}{15} & \progressbarTimeout{0}{22}{0} & \progressbarA{6}{22}{6} & \progressbarFN{14}{22}{14} & \progressbarTimeout{2}{22}{2} & \progressbarA{0}{22}{0} & \progressbarFN{22}{22}{22} & \progressbarTimeout{0}{22}{0} & \progressbarA{14}{22}{14} & \progressbarFN{8}{22}{8} & \progressbarTimeout{0}{22}{0} & \progressbarA{20}{22}{20} & \progressbarFN{2}{22}{2} & \progressbarTimeout{0}{22}{0} \\
  \textbf{RE} (16) & \progressbarA{8}{16}{8} & \progressbarFN{8}{16}{8} & \progressbarTimeout{0}{16}{0} & \progressbarA{10}{16}{10} & \progressbarFN{0}{16}{0} & \progressbarTimeout{6}{16}{6} & \progressbarA{10}{16}{10} & \progressbarFN{6}{16}{6} & \progressbarTimeout{0}{16}{0} & \progressbarA{11}{16}{11} & \progressbarFN{3}{16}{3} & \progressbarTimeout{2}{16}{2} & \progressbarA{4}{16}{4} & \progressbarFN{12}{16}{12} & \progressbarTimeout{0}{16}{0} & \progressbarA{16}{16}{16} & \progressbarFN{0}{16}{0} & \progressbarTimeout{0}{16}{0} & \progressbarA{16}{16}{16} & \progressbarFN{0}{16}{0} & \progressbarTimeout{0}{16}{0} \\
  \textbf{TO} (2) & \multicolumn{3}{c|}{n/a} & \progressbarA{2}{2}{2} & \progressbarFN{0}{2}{0} & \progressbarTimeout{0}{2}{0} & \multicolumn{3}{c|}{n/a} & \multicolumn{3}{c|}{n/a} & \multicolumn{3}{c|}{n/a} & \progressbarA{2}{2}{2} & \progressbarFN{0}{2}{0} & \progressbarTimeout{0}{2}{0} & \progressbarA{2}{2}{2} & \progressbarFN{0}{2}{0} & \progressbarTimeout{0}{2}{0} \\
  \midrule
  All (108)& \hspace*{0.06cm}\progressbarA{13}{108}{13} & \progressbarFN{22}{108}{22} & \progressbarTimeout{1}{108}{1} & \hspace*{0.06cm}\progressbarA{35}{108}{35} & \progressbarFN{14}{108}{14} & \progressbarTimeout{37}{108}{37} & \hspace*{0.06cm}\progressbarA{49}{108}{49} & \progressbarFN{57}{108}{57} & \progressbarTimeout{0}{108}{0} &  \hspace*{0.06cm}\progressbarA{53}{108}{53} & \progressbarFN{41}{108}{41} & \progressbarTimeout{12}{108}{12} & \hspace*{0.06cm}\progressbarA{36}{108}{36} & \progressbarFN{70}{108}{70} & \progressbarTimeout{0}{108}{0} & \hspace*{0.06cm}\progressbarA{91}{108}{91} & \progressbarFN{17}{108}{17} & \progressbarTimeout{0}{108}{0} & \hspace*{0.2cm}\bfseries \progressbarA{105}{108}{105} & \bfseries \progressbarFN{3}{108}{3} & \bfseries \progressbarTimeout{0}{108}{0} \\
  \bottomrule
  \end{tabular}} }\\

  \resizebox{1.0\textwidth}{!}{ 
  \begin{tabular}{rclcclllccc}
      \midrule
      \multicolumn{5}{l}{\textit{Real-World Smart Contracts: Individual Solidity Files}\quad\faHandOUp} &  \multicolumn{5}{r}{\faHandODown \quad\textit{Real-World Smart Contracts: Whole DApp Projects}} \\ \midrule

      \bfseries DApp Projects & \bfseries Num. & \bfseries Source & \bfseries Trig. &\bfseries Severity & \bfseries Location & \bfseries Vul. & \bfseries \#R & \bfseries  \#Fail & \bfseries Time \\ \midrule
        SkaleToken&  49 & \url{https://github.com/skalenetwork} & \ding{51} & High  &\tabincell{l}{SkaleBalances.sol\\L55-L68}  & \texttt{RE} &4 & 7 &  \SI{3}{m} \SI{37.254}{s/c}\\  
      \midrule
        Gamma &  17 & \url{https://github.com/GammaStrategies}  & \ding{51} & High & \tabincell{l}{UniProxy.sol \\ L75-L82} & \texttt{RE}& 2 & 0 &  \SI{1}{m} \SI{54.471}{s/c}\\
      \midrule
        OminiBridge&  45 & \url{https://github.com/omni/omnibridge}  & \ding{51} & High & \tabincell{l}{TokensRelayer.sol\\L99-L117} & \texttt{RE} & 4 & 9 &  \SI{3}{m} \SI{13.069}{s/c}\\   \midrule
        0x\_Exchange& 56 & \url{https://github.com/0xProject}  & \ding{51} & High & \tabincell{l}{Meta...Feature.sol\\ Lines:98-106} & \texttt{RE}& 5 & 11 &  \SI{3}{m} \SI{44.082}{s/c}\\
        \midrule
        SuperLauncher &  14 & \url{https://github.com/SuperLauncher}  & \ding{51}& High & \tabincell{l}{MainWorkflow.sol \\ L340-L351} & \texttt{RE}& 3 & 0 &  \SI{1}{m} \SI{48.326}{s/c}\\
      \midrule
        AloeBlend&  40 & \url{https://github.com/aloelabs} & \ding{51} & Medium  & \tabincell{l}{Factory.sol\\L54-L67}  & \texttt{TO} &4 & 5 &  \SI{2}{m} \SI{30.226}{s/c}\\  
      \midrule
        PoolTogether&  75 & \url{https://github.com/pooltogether} & \ding{51} & Medium  & \tabincell{l}{TokenFaucet.sol\\L119-L138}  & \texttt{BD} & 5 & 16 &  \SI{4}{m} \SI{12.204}{s/c}\\  
      \midrule
        MolochDAO&  13 & \url{https://github.com/MolochVentures} & \ding{51} & High  & \tabincell{l}{MolochD.sol\\L590-L599}  & \texttt{EL} &5 & 1 &  \SI{2}{m} \SI{47.347}{s/c}\\  
      \midrule
        tBTC&  47 & \url{https://github.com/threshold-network} & \ding{51} & High  & \tabincell{l}{MolochD.sol\\L590-L599}  & \texttt{EL} &7 & 4 &  \SI{5}{m} \SI{07.194}{s/c}\\  
      \midrule


      1inch Liquidity$^h$&  66 & \url{https://github.com/1inch}  & \ding{51} & High & \tabincell{l}{Refer...Receiver.sol\\L38-L50} & \texttt{EL} & 6 & 13 &  \SI{4}{m} \SI{27.492}{s/c}\\ 
        \bottomrule        
  \end{tabular}}
\end{table*}

\subsection{Performance on Real-world Contracts (RQ3)}
\label{sec:realworld}
As shown in Table 3, we further evaluate SmartFuzz on real-world smart contracts, including both individual Solidity files and  DApp projects, to assess its performance across a broader spectrum of vulnerability types. Specifically, we also benchmark SmartFuzz against state-of-the-art detectors on the individual contracts from D2, including symbolic executors (e.g., Oyente, Mythril) and fuzzers (e.g., sFuzz, ConFuzzius, Smartian, and MuFuzz). The baselines are not strictly identical to those used before, mainly because certain earlier tools (e.g., SmarTest) are limited to a small set of vulnerability types.
Compared with the previous experiment, which used 10 randomly selected D1 contracts and their transaction sequences as learning examples, all real-world contracts from D2 are treated as unseen contracts. For a fair comparison, we adopt the same evaluation metrics as Mufuzz and report the results of each tool in terms of true positives and false negatives. Moreover, we also count the number of timeout and error cases reported by each tool.

\noindent\textbf{Individual Solidity Files.} 
Table 3 (upper part) presents a performance comparison of SmartFuzz and baseline tools across six vulnerability types. Overall, SmartFuzz identified $97.2\%$ ($\frac{105}{108}$) true vulnerabilities with only 3 false negatives. Specifically, it outperforms all the baselines by 1.15x to 8x, with the range derived from $\frac{105-91}{91}$ for MuFuzz  to $\frac{105-13}{13}$ for Oyente, and fully supports all bug classes.
Compared to the strongest baseline MuFuzz, , SmartFuzz achieves 0 FN for UE and reduces FNs by 80\% ($\frac{5-1}{5}$) for BD and 75\% ($\frac{8-2}{8}$) for EF. This is because MuFuzz depends solely on data-flow-based feedback to construct static transaction ordering and, in failure situations, hardly performs global or local sequence adjustments to handle runtime execution exceptions. By contrast, SmartFuzz’s CRP and multi-agent team employ real-time environmental feedback to iteratively refine the entire sequences.

\noindent\textbf{DApp Projects.} 
Furthermore, we implement SmartFuzz to analyze more than 100 DApps in total and then conduct a manual audit for each project. 
These DApp projects  are considered much more diverse and complex than regular contracts due to their larger codebase and more complex transaction handling \cite{zheng2024dappscan}. 
Table 3 (lower part) shows 10 DApp projects with the most severe vulnerabilities, along with their locations, severity levels, and reflection rounds. Such listed vulnerabilities could cause substantial economic losses, highlighting the practical effectiveness of SmartFuzz and its potential real-world impact.
Specifically, in these DApp projects, an average of 12.36\% of the analyzed contract files failed, with most failures resulting from exceeding the time limit (10 minutes) or the maximum number of reflection rounds.
On average, each vulnerable contract underwent 4.5 reflection rounds, and the analysis time ranged from 1 to 5 minutes. The longest case occurred in the \texttt{tBTC}, which took 7 rounds to detect an ether-leaking vulnerability, mainly because each file contained a large amount of code. Furthermore, for each audited DApp project, SmartFuzz automatically reports the severity level, and the file, function, and line involved in each vulnerability, providing developers with precise guidance for locating and fixing these bugs in real-world scenarios.
Moreover, beyond prior approaches including SmarTest, ILF, and RLF, SmartFuzz benefits from its ability to treat an entire DApp project as a unified input and consider dependencies between the interacting contracts, which allows it to detect vulnerabilities that require analyzing interactions across multiple contracts to reveal flaws (e.g., reentrancy).

\answerbox{
\textbf{Answer to RQ3:} On real-world contracts, SmartFuzz detects 97.2\% of true vulnerabilities, outperforming baselines by 1.15x to 8x and reducing false negatives by up to 80\%. Besides, it successfully detects complex vulnerabilities in all DApp projects  with an average of 4.5 reflection rounds per  contract.
}

\section{Related Work} 
\label{sec:related}

\noindent\textbf{Smart Contract Vulnerability Detection.} The detection of smart contract vulnerabilities can  be roughly  divided into dynamic analysis  and static analysis,  depending on whether the programs need to be executed.
Within static analysis, formal verification validates the logical integrity of smart contracts, such as theorem proving (e.g., ConCert \cite{annenkov2020concert}, Isabelle/hol \cite{amani2018towards}), model checking (e.g., Zeus \cite{DBLP:conf/ndss/KalraGDS18}).
Symbolic execution, pioneered by Oyente \cite{luu2016making}, detects vulnerabilities by exploring feasible execution paths and has been  leveraged by later tools such as  Mythril \cite{mueller2017mythril}, Manticore \cite{mossberg2019manticore},  DefectChecker \cite{chen2021defectchecker}, Sailfish \cite{bose2022sailfish} and so on.
 Moreover, another critical branch of static analysis involves the semantic and syntactic analysis of source code. Customized intermediate representations are designed to facilitate a better understanding of the logic and structure of programs, and have been implemented in tools such as SmartCheck \cite{tikhomirov2018smartcheck}, EthIR \cite{albert2018ethir}, ClairvOyance \cite{ye2020clairvoyance}, and Slither \cite{feist2019slither}.
Several recent studies \cite{zhuang2021smart, wu2021peculiar, zhang2022reentrancy, sendner2023smarter} design various deep learning-based models to detect vulnerabilities by  either treating the source code as a text sequence or casting the rich control and data flow semantics of the source code into graphs for the learning process.
By contrast, dynamic analysis generally involves simulating program execution and observing its dynamic behavior to uncover vulnerabilities, typically through fuzzing tests.
For example,
 ReGuard \cite{liu2018reguard} is a fuzzing-based analyzer designed to systematically detect reentrancy vulnerabilities in  contracts by generating a series of random but strategically varied transactions. ContractFuzzer \cite{jiang2018contractfuzzer} detects vulnerabilities by generating fuzzing inputs based on the ABI specifications. sFuzz \cite{nguyen2020sfuzz} employs an efficient, lightweight multi-objective adaptive strategy to select seeds for the fuzzing process.
 These fuzzers primarily focus on generating specific inputs to satisfy complex conditions while failing to consider the transaction sequences to trigger vulnerabilities.

\noindent\textbf{Machine Learning-guided  Fuzzing.} 
Machine learning techniques have recently been explored to improve fuzzing process. For example, generative models trained on existing fuzzing input datasets have been developed in several works \cite{godefroid2017learn, wang2017skyfire, cummins2018compiler}.
 AFLFast \cite{bohme2016coverage} and Neuzz \cite{she2019neuzz} further model program branching behavior using Markov Chains and neural networks, respectively, trained on fuzzing-generated inputs. 
Furthermore, CovRL-Fuzz \cite{eom2024fuzzing} employs large language models combined with reinforcement learning to optimize the fuzzing process for identifying vulnerabilities in JavaScript interpreters. FuzzGPT \cite{deng2024large} leverages large language models to generate unusual programs for fuzzing deep learning libraries by learning from historically bug-triggering code snippets.
In the field of smart contract vulnerability detection, to generate vulnerable transaction sequences,  ILF \cite{he2019learning} leverages imitation learning to generate vulnerable transaction sequences by learning from symbolic execution experts during fuzzing. 
RLF \cite{DBLP:conf/kbse/SuDZZL22} employs reinforcement learning with rewards derived from identified vulnerabilities and code coverage to steer transaction sequence generation. 
However, these methods fail to understand program semantics in advance, resulting in generating numerous invalid transaction sequences and wasting  considerable resources on code regions unrelated to vulnerabilities.

\section{Conclusion}

In this work, we present a new  Collaborative Reflection fuzzer, namely \texttt{SmartFuzz}, for effectively generating vulnerable transaction sequences to uncover the underlying vulnerabilities in smart contracts. In particular, we propose a new continuous reflection process for fuzzing smart contracts, which models the transaction sequence generation as a self-evolving   process. Besides, we design a reactive collaborative chain to decompose the fuzzing process into multiple subtasks and arrange the execution order of each agent. The RCC is designed to facilitate real-time interaction with the external environment. Furthermore, we design a multi-agent collaborative team that implements a hierarchical reflection strategy to consider both local and global dependency within a transaction sequence. This team operates along the RCC, enabling it to continuously generate and refine vulnerable transaction sequences using real-world feedback.
Extensive experimental results  demonstrate that SmartFuzz significantly outperforms existing tools across real-world smart contracts and  DApp projects.

\bibliographystyle{IEEEtran}
\bibliography{IEEEabrv_full,SmartFuzzer_ref}

\newpage

\end{document}